\documentclass{aa}
\usepackage{graphics}

\begin{document}

   \thesaurus{06.08.14.1, 06.08.16.7, 06.08.18.1, 06.08.15.1}

   \title{Crab Pulsar Photometry and the Signature of Free Precession}

   \author{Andrej \v Cade\v z, Simon Vidrih, Mirjam Gali\v ci\v c
          \inst{1}
          \and
          Alberto Carrami\~ nana\inst{2}
          }

   \offprints{A. \v Cade\v z}

   \institute{Department of Physics, FMF University in Ljubljana,
              Jadranska 19, 1000 Ljubljana, Slovenia\\
              email: andrej.cadez@uni-lj.si
         \and
             I.N.A.O.E., Luis Enrique Erro 1, Tonantzintla\\
              Puebla 72840, M\'exico
             }

   \date{Received 03 July 2000 / Accepted 16 November 2000}

	\authorrunning{\v Cade\v z et al.}
   
	\maketitle

   \begin{abstract}

   Optical photometry for the pulsar PSR0531+21 has been extended with new      
	observations that strengthen evidence for a previously observed 60 seconds 
   periodicity. This period is found to be increasing with time at approximately
	the same rate as the rotational period of the pulsar. The observed period    
	and its time dependence fit a simple free precession model. 

      \keywords{Stars: neutron -- pulsars: individual: PSR0531+21 --
                rotation --
                oscillations 
               }
   \end{abstract}

%________________________________________________________________

\section{Introduction}

   More than seven years ago we began an investigation of the optical light     
	curve of the Crab pulsar. We tentatively identified a peak in Fourier spectra
   of the light curve at a frequency $\approx 1/60\,$Hz (\v Cade\v z \& 
   Gali\v ci\v c~\cite{c4} a,~\cite{c5} b,~\cite{c6} c). Later observations     
	slightly enhanced the signal to noise ratio of the suggested peak. We        
	suspected that it was the signature of free precession of the pulsar and thus
	in~\cite{c7} proposed a simple model. We tested it by predicting the free    
	precession frequency of the Earth and showed that it also reproduces well the
	measured ellipticities of the planets (\v Cade\v z et al \cite{c7}). The     
	former test was based on the argument that the relative rigidity of the      
	pulsar is at least as strong as that of the Earth. This allowed us to apply  
	the same model to calculate the free precession frequency of pulsars. We     
	found that the observed $1/60\,$Hz frequency is consistent with a            
	$1.3\,$M$_\odot$ pulsar model based on the tensor interaction equation of    
	state. As a further check we proposed to measure the slow-down rate of free  
	precession. In the past years we gathered more data, building a set of more  
	than 3.400 images, which cover more than 20 hours of photometry spanning a   
	period of almost nine years. It is appropriate to confront these data with   
	the proposed test. 

%__________________________________________________________________

\section{Calibrations and data sets}

\label{Tabela1}
\begin{table}[b]
\begin{centering}
\caption{List of data sets giving the {\it name} of the run (as used in later
references), the name of the {\it telescope}, the {\it date} of observation,    
the {\it sampling rate} (the exposure time to obtain the sample is generally    
shorter than the sampling rate) and the total {\it duration} of the run.}
\begin{tabular}{|c|c|c|c|c|}
\hline Name & Telescope & Date & S.Rate [s]& Dur. [s]\\
\hline
\hline Oct.15.91 & HST & Oct. 15. 91 & 4.0 & 1770\\
\hline Dec.12.94 & CE & Dec. 12. 94 & 22.7 & 3540\\
\hline Dec.19.95 & CE & Dec. 19. 95 & 26.7 & 5420\\
\hline Dec.17.96 & CE & Dec. 17. 96 & 22.1 & 2120\\
\hline Feb.22.97 & As & Feb. 22. 97 & 24.9 & 3760\\
\hline Feb.26.97a & As& Feb. 26. 97 & 24.4 & 3780\\
\hline Feb.26.97b & As & Feb. 26. 97 & 24.4 & 2930\\
\hline Dec.6.97 & As & Dec. 6. 97 & 26.7 & 2960\\
\hline Dec.7.97a & As & Dec. 7. 97 & 26.7 & 2670\\
\hline Dec.7.97b & As & Dec. 7. 97 & 26.7 & 4801\\
\hline Dec.7.97c & As & Dec. 7. 97 & 26.7 & 3390\\
\hline Jan.22.98 & GH& Jan. 22. 98 & 25.4 & 9870\\
\hline Jan.23.98 & GH & Jan. 23. 98 & 22.4 & 4140\\
\hline Jan.24.98 & GH & Jan. 24. 98 & 23.8 & 9460\\
\hline Jan.25.98 & GH & Jan. 25. 98 & 26.1 & 4640\\
\hline Oct.21.99 & GH & Oct. 21. 99 & 19.9 & 8720\\
\hline
\end{tabular}
\end{centering}
\end{table}

Our photometry was done in the stroboscopic mode. This type of photometry was   
introduced previously (\v Cade\v z \& Gali\v ci\v c \cite {c6}) in order to     
increase the signal to noise ratio of measured   pulsed emission from pulsars.  
The method is based on the fact that the arrival time and duration of pulses    
can be calculated quite precisely. Thus we introduced the stroboscope - a phase 
controlled shutter (rotating wheel), which passes light to the CCD detector only
during the time intervals when the pulsar (main pulse) is ``on''. Therefore,      
images obtained behind this shutter receive all light emitted by the pulsar, but
the much brighter emission from the surrounding nebula, is reduced by the ratio 
of the expected main pulse duration to the pulse period (set to 0.1 by our      
stroboscope). In our case this noise reduction ratio is about a factor of 2. 
In stroboscopic photometry the timing is crucial, so we spent a considerable    
effort checking it. The whole stroboscopic setup together with the CCD has been 
periodically tested at the Hewlett-Packard calibration laboratory in Ljubljana, 
where an artificial pulsar has been pulsed by a cesium clock~(Gali\v ci\v c     
\cite{c9}). All tests confirmed that the stroboscopic wheel correctly follows   
the frequency and phase of the input signal\footnote{During the observations    
abbreviated by $As$ we had some timing problems which were traced to a software 
bug in calculating the proper date, so that the calculated barycentric frequency
set on the timer was slightly off, resulting in slow phase slippage of the      
stroboscope with respect to the pulsar. The slippage was slow enough that we    
could calculate its effect and correct for it. }. 

\begin{figure*}
\begin{centering}
  \resizebox{13.5cm}{!}{\includegraphics{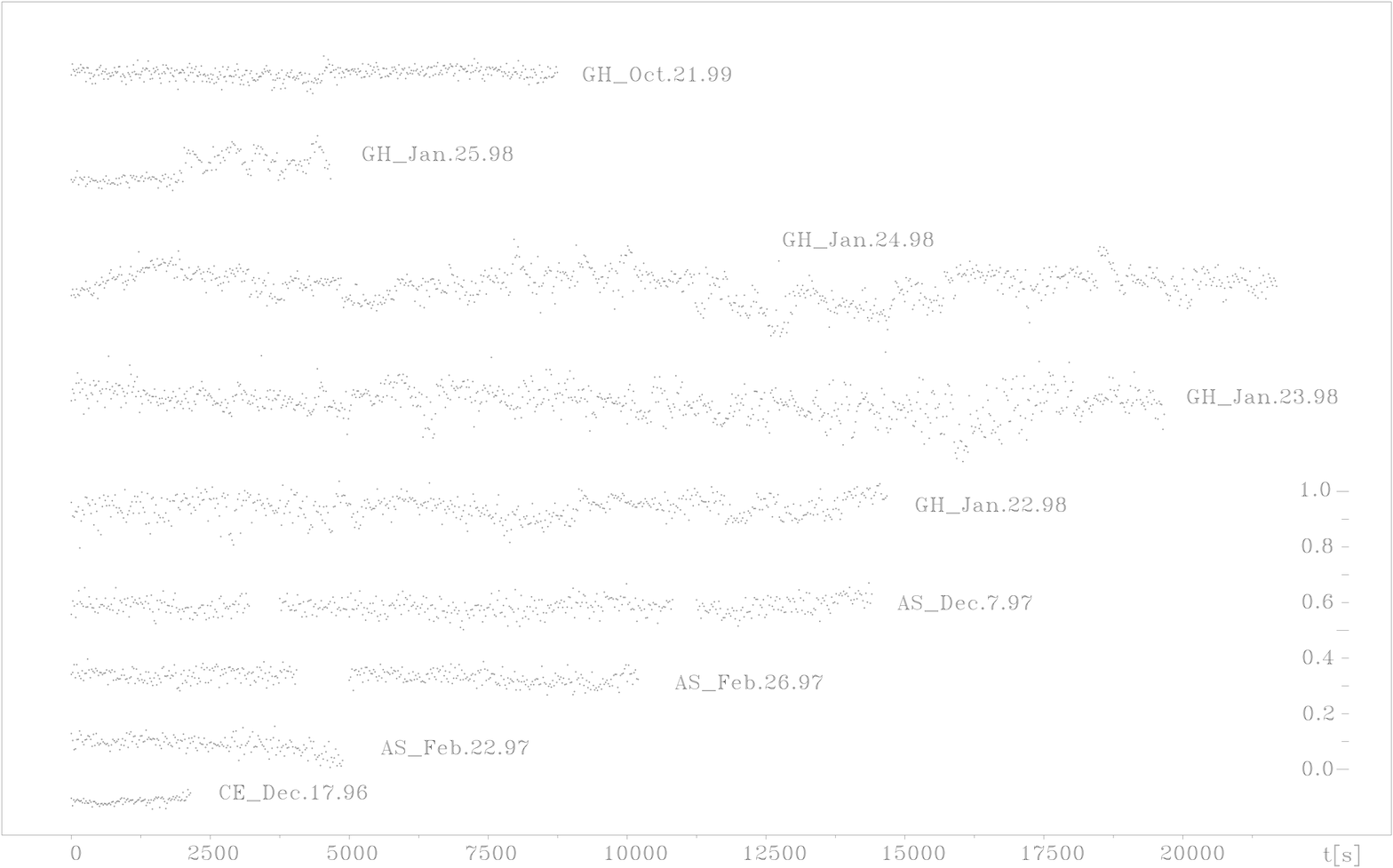}}
  \hfill
  \parbox[b]{55mm}{\caption[Fig2]{Light curves of the comparison star.}}
\label{fig:Fig1}
\end{centering}
\end{figure*} 

\begin{figure}
  \resizebox{\hsize}{!}{\includegraphics{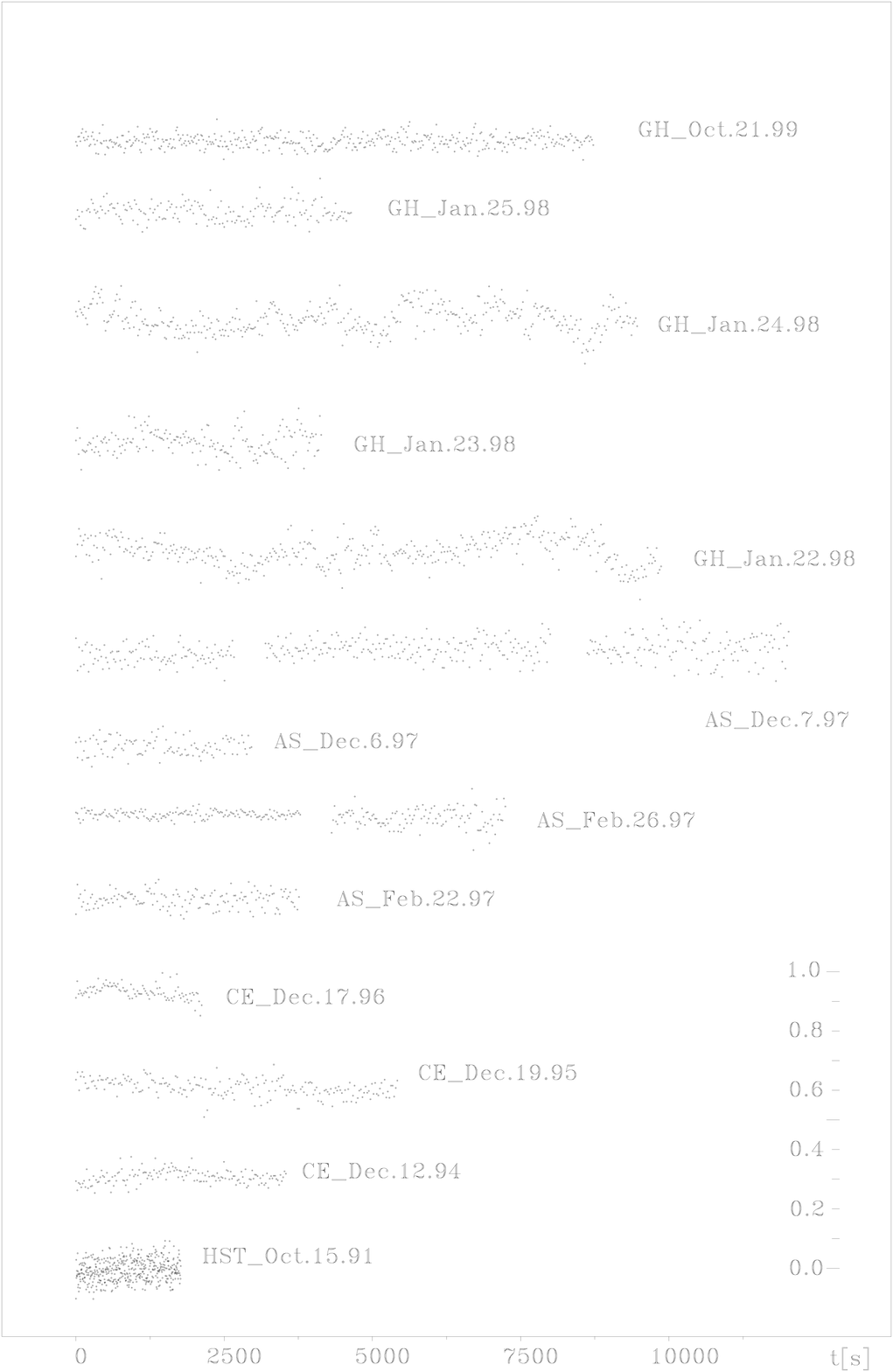}}
  \caption{All pulsar light curves. The relative magnitude scale is at lower    
  right. }
  \label{fig:Fig2}
\end{figure} 

Photometric data were taken with four different telescopes and exposure times
between 4 and 15 seconds\footnote{The 4 seconds sampling rate of the HST data is
not intrinsic to the high speed photometer but rather to our processing         
procedure.} and sampling rates beween 4.0 and 26.7 seconds. The data are of a   
somewhat varying quality depending on the telescope and on respective observing
conditions. In Table 1 we list the dates, telescopes, durations of observation  
and sampling rates of all runs. The abbreviations  HST (Percival et al.         
\cite{c13})  CE, As and  GH stand for Hubble space telescope, the 1.82 m        
telescope of the Asiago observatory, the 1.22 m telescope of the Asiago         
observatory and the 2.12m telescope of the Guillermo Haro observatory           
respectively.

In Fig. 1 and 2 we show the light curves of a field star, which is used for     
calibration and the contemporeneous light curves of the Crab pulsar. The        
photometry was done by using IRAF/DAOPHOT and previously DAOPHOT                
routines\footnote{Until 1998 the DOS version of DAOPHOT available at            
www.fiz.uni-lj.si/astro/daophot.html and later IRAF V2.11 has been used.}. 

%______________________________________________________________

\section{Data analysis and results}

In order to search for a possible common modulation frequency in all Crab light 
curves, we construct a matrix with components $M_{jk}(\omega)$, which are       
Fourier transforms of cross-correlation functions between light curves, i.e.: 
\begin{equation}
M_{jk}(\omega)=\vert F_j(\omega) F_k(\omega)\vert~~~,
\end{equation}
and $F_k(\omega)$ is the Fourier transform of the $k^{th}$ light curve:
\begin{equation}
F_k(\omega)={1\over \sqrt{2 \pi T_k}}\sum_l m_k(t_l)~e^{i\omega t_l}~           
{{t_{l+1}-t_{l-1}}\over 2}~~~.
\end{equation}
Here $m_k(t_l)$ is the magnitude as measured from the $l^{th}$ image of the     
k$^{th}$ light curve taken at time $t_l$ (with respect to the beginning of the  
$k^{th}$ set) and $T_k=t_{l_{max}}-t_{l_{min}}$ is the total duration of the    
k$^{th}$ set of measurements. $F_k(\omega)$ is considered as a continuous       
function defined first at discrete frequencies $\omega_s={2\pi \over T_k} s$,   
where s is an integer belonging to the interval $0\leq s \leq ({l_{max}}-       
l_{min})/2$, and then extended so that $\vert F_k(\omega) \vert=                
\vert F_k(\omega_s) \vert$ on the interval ${2\pi \over T_k} (s-{1\over 2})     
\leq\omega < {2\pi \over T_k} (s+{1\over 2})$.

We first test statistical properties of the $M_{jk}$ components on light curves 
of the field star, i.e. we test the hypothesis that all noise in the light curve
is white Gaussian noise. If this hypothesis is correct, then (for each $\omega$)
$M_{ij}$ ($i\neq j$) may be considered as a realization of a random process     
$\mu$ distributed according to: 
\begin{equation}
W(\mu)= {\mu \over \sigma^4}K_0\left({\mu \over \sigma^2}\right )~~~,
\end{equation}
and $M_{ii}$ belongs to a random process $\nu$ distributed according to
\begin{equation}
U(\nu)={1\over 2 \sigma^2} e^{-\nu/(2\sigma^2)}~~~,
\end{equation}
where\footnote{The average is taken over all k and $\omega $ and, since noise is
considered as white, all $F_k$ belong to the same random process.} $\sigma ^2   
=\langle \vert F_k(\omega)\vert ^2\rangle$ is the spectral density of noise in  
measuring the magnitude, $K_0(x)$ is the modified Bessel function               
(Gali\v ci\v c \cite{c9}) and $\langle \rangle$ indicates the average over      
frequencies (higher than 0.005Hz to exclude the 1/f noise at the low end). In   
order to detect a periodic component in the light curve we construct two        
quantities\footnote{In fact, since the light curves are of varying quality, we  
define $\overline M$ and $\sigma_M$ as weighted averages, where the weight of   
each component $M_{jk}(\omega)$ is $1\over \sigma_j^2 \sigma_k^2$, where        
$\sigma_{j(k)}$ is the average  of $\langle \sigma(\omega)\rangle $             
corresponding to the $j^{th}$($k^{th}$) light curve.}:
 \begin{equation}
 \overline M(\omega)={1\over N^2}\sum_{i=1}^N\sum_{j=1}^N M_{ij}(\omega)
 \end{equation}
 and
 \begin{equation}
 \sigma_M(\omega)=\sqrt{{1\over N^2}\sum_{i=1}^N\sum_{j=1}^N M_{ij}^2(\omega)~-~{\overline M}^2(\omega)}~~~,
 \end{equation} 
 which in case of white Gaussian noise average to:
 \begin{equation}
 \label{povpM}
 \langle \overline M(\omega )\rangle={\pi N(N-1)+ 8 N\over 2 N(N+1)}\sigma^2
 \end{equation}
 and:
 \begin{eqnarray}
 \label{sigmaM}
 \lefteqn{ \langle \sigma_{M}(\omega )\rangle ={}}
 \\& & {}
 {\sqrt{(16-\pi^2)N^2 + 2(32+\pi^2 - 8\pi)N +(16\pi -16-\pi^2)}\over 2 (N+1)}   
 \sigma^2{}.
 \nonumber
 \end{eqnarray}
 
 \begin{figure}
   \resizebox{\hsize}{!}{\includegraphics{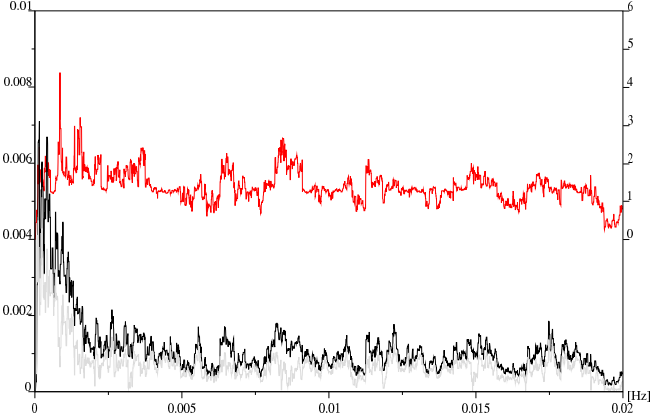}}
   \hfill
   \parbox[b]{55mm}{\caption[Fig3]{Field star: average cross-correlation        
	function (middle curve), its sigma (lower grey curve) and the median ratio of
	the two (the upper curve) as the function of frequency ($\nu = {\omega \over 
	2 \pi}$). The LHS scale is in ${mag^2\over Hz}$ and relates to the bottom    
	curves and the RHS scale relates to the ratio.}}
  \label{fig:Fig3}
 \end{figure} 
 
In Fig. 3 we plot $\overline M(\omega)$, $\sigma_M(\omega)$ and $\overline      
M(\omega)\over \sigma_M^\prime(\omega)$ calculated from the 12 light curves of  
the field star as plotted in Fig. 2. Here $\sigma_M^\prime(\omega)=             
\sigma_M(\omega)$ if $\sigma_M(\omega)> \langle \sigma_M(\omega)\rangle $ and   
$\sigma_M^\prime(\omega)=\langle \sigma_M(\omega)\rangle $ if                   
$\sigma_M(\omega)\le \langle \sigma_M(\omega)\rangle $. The average $\langle    
\overline M(\omega)\rangle $ = 0.00091${mag^2\over Hz}$ and $\langle            
\sigma_M(\omega)\rangle $ = 0.00054${mag^2\over Hz}$. Thus, from eq.\ref{povpM} 
we deduce  $\sigma=0.023{mag\over \sqrt{Hz}}$ and eq.\ref{sigmaM} gives $\sigma=
0.020{mag\over \sqrt{Hz}}$. Both numbers are consistent with each other and with
average dispersion estimates from IRAF ($\sigma_{IRAF}=\sigma/\sqrt{\tau};~ \tau
={\it exposure~ time}$). Therefore, we conclude that the test star data conform 
with the statistical assumptions described above.
 
\begin{figure}
   \resizebox{\hsize}{!}{\includegraphics{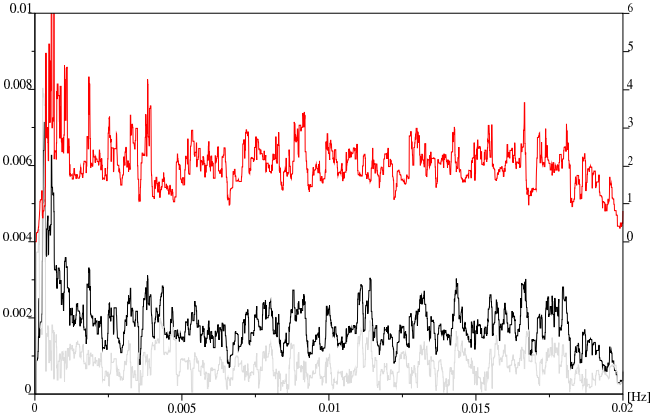}}
   \hfill
   \parbox[b]{55mm}{\caption[Fig4]{The average cross-correlation function, its  
	sigma and the median ratio for pulsar data. Symbols used are the same as     
	those in Fig. 3.}}
   \label{fig:Fig4}
 \end{figure} 

The result of a similar analysis for 16 Crab light curves is shown in Fig. 4. It
is apparent that $\langle {\sigma_M}(\omega)\rangle $ is almost the same for
the Crab and the test star; however the average $\langle \overline
M(\omega)\rangle $ for the Crab is 1.5 times that of the test star. The process 
$M_{ij}(\omega)$ can, therefore, not be considered as white Gaussian noise, but 
it can be understood by assuming that it is the sum of a stationary ``signal''  
$M^0_{ij}(\omega)$ and white Gaussian noise $M^n_{ij}(\omega)$ with the same    
spectral density as that of the test star. We considered the possibility that   
the ``signal'' is connected to a spurious phase modulation in stroboscope       
timing, but lab tests have set the upper limit for such a modulation some 100   
times lower~(Gali\v ci\v c \cite{c9}). 
We therefore conclude that the Crab pulsar light curve is intrinsically         
intensity modulated at the level of about 0.03$mag\over \sqrt{Hz}$. The signal  
to noise ratio of this modulation measure is presently too low to reveal more   
about its nature. 
 
To test for the slow down of the purported free precession we recalibrate       
frequencies in $F_k(\omega)$ to the same date according to an assumed power law 
dependence of the free precession frequency with time, i.e. we introduce the    
recalibrated frequency $\tilde \omega=\omega ({\nu_{ref}\over \nu_k})^p$, where 
$\nu_{ref}$ and $\nu_k$ are the rotation frequencies of the pulsar at the       
reference date (Jan.1.1996) and at the date when the k$^{th}$ light curve was   
obtained. The above analysis has been repeated for different powers $p$ between 
0 and 4. A 5.3 $\sigma$ peak stands out in $\overline M(\tilde \omega)/         
\sigma_M^\prime(\tilde\omega)$ and is found at $\tilde \omega =                 
{2 \pi \over 59.93 s}$ for $p$ around 1, as shown in Fig.5 and 6. 
Note that in the free precession model $p$ is expected to be 1 if the time for  
the pulsar to relax its internal stress is much longer than the total           
observation period ($\approx 9$years). It is expected to be 3, if this          
relaxation time is considerably shorter. Short relaxation times would point to  
externally driven precession.

\begin{figure}
   \resizebox{\hsize}{!}{\includegraphics{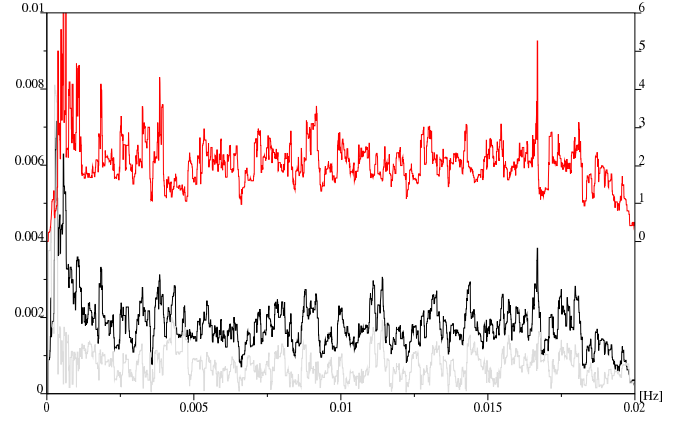}}
   \hfill
   \parbox[b]{55mm}{\caption[Fig5]{The average cross-correlation function ($\bar
	M(\tilde \omega)$), its sigma ($\sigma_M(\tilde \omega)$) and the median
	ratio $\left( {{\bar M(\tilde \omega)}\over {\sigma^\prime_M(\tilde
	\omega)}}\right)$ for pulsar data as the function of the recalibrated
	frequency  $\tilde \nu={\tilde \omega \over 2
	\pi}={\omega \over 2 \pi}\left ({\nu_{ref}\over \nu_k}\right
	)$, (i.e. $p=1$). Other symbols are the same as in Fig 3.}}
  \label{fig:Fig5}
\end{figure} 

\begin{figure}
   \resizebox{\hsize}{!}{\includegraphics{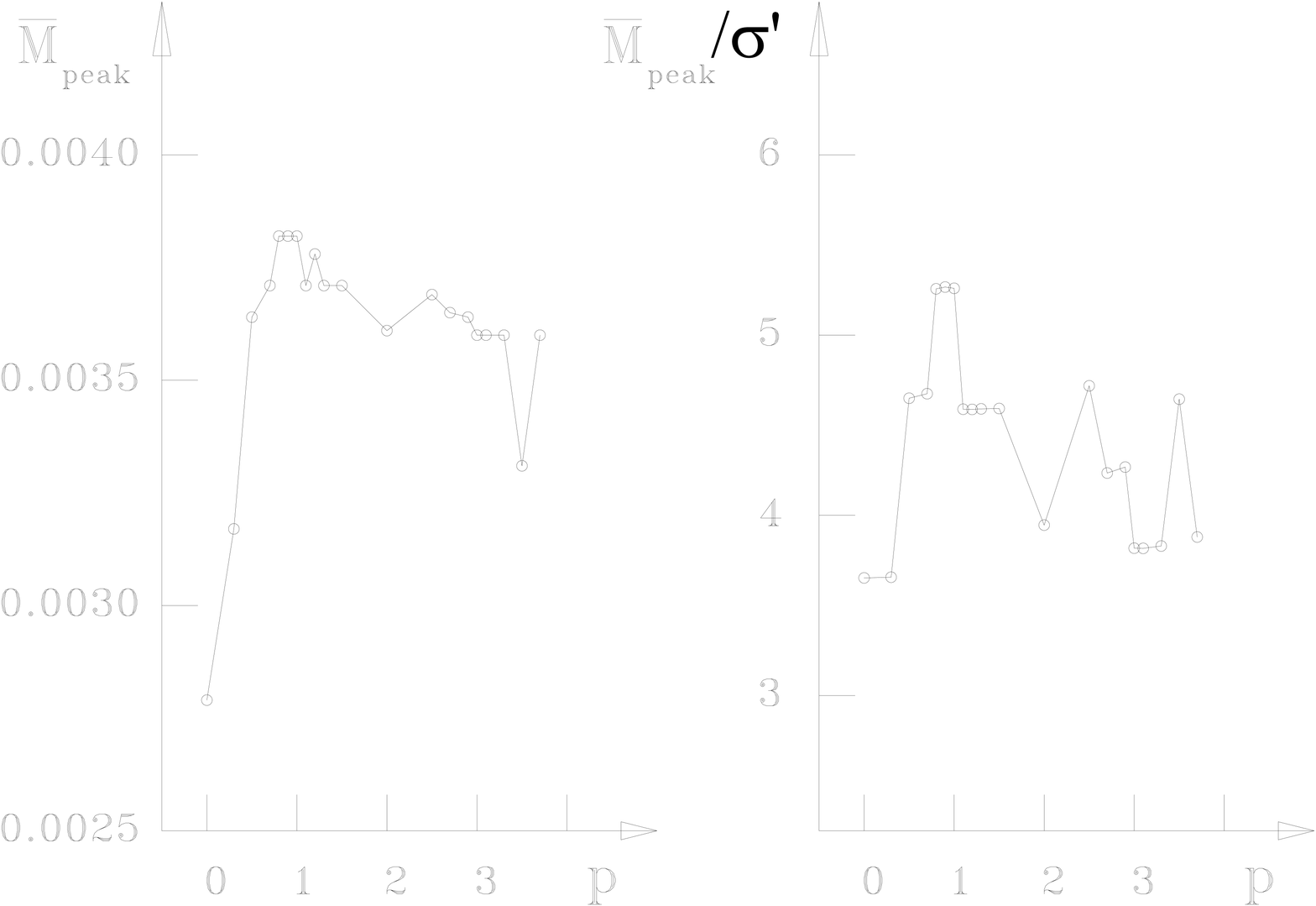}}
   \hfill
   \parbox[b]{55mm}{\caption[Fig6]{The peak $\overline M(\tilde \omega)$ and    
	$\overline M(\tilde \omega)/\sigma^{\prime}(\tilde \omega)$ as function of   
	the power law index $p$}}
  \label{fig:Fig6}
\end{figure} 

%_______________________________________________________________________________

\section{Discussion}
The question whether pulsars free precess was raised soon after pulsars
were identified as rotating neutron stars. However, more than twenty years
later most astronomers would agree with Trimble and McFadden's (\cite{c19})
somewhat humorous reference to our paper suggesting that Earth is
the only known free precessing body in the Universe. We believe that pulsar free
precession has evaded detection and/or identification for two main reasons.    
1) Most researchers assume that neutron star matter is in a complex superfluid 
state, which provides an excuse to evade the question: what is the inertial     
eccentricity of a rotating neutron star? As a result, the range
of predicted or ``identified'' free precession frequencies in the literature is
almost without bounds, although Pines and Shaham (\cite{c16}) estimated 
 that the Crab pulsar would wobble with a period $\simeq$ 5 minutes if it
had a solid core. 2) As Alpar and \"Ogelman (\cite{c1}) correctly point out,    
free precession is a motion characteristic of rigid rotators. In a plastic      
rotator, free precessing motion is damped out by viscous losses - in pulsars    
this should happen on a time scale of years or thousands of years. Therefore,   
unless pulsars are exposed to reasonably large external torques, they are not   
expected to free precess. 
In our earlier paper (\cite{c7}) we tried a naive approach: starting with the   
observation that the Earth free precesses but taking into account the           
possibility that it may not be quite rigid with respect to free precession. We  
analysed the relation between the Earth's inertial eccentricity, free precession
frequency, rotation frequency and obliquity. We found that the Earth's shape and
that of other planets is quite similar to the equilibrium shape of
rotating polytropes with no shear stress. The Earth's free precession
frequency can be fairly accurately calculated on the basis of these assumptions.
We argue that a pulsar's crust is relatively more rigid and thick than the
Earth's. Therefore, argument for planets can also be used to calculate
a pulsar's free precession frequency. For the fast Crab pulsar we found a free
precession frequency  on the order of $minutes^{-1}$ and in particular the
theoretical value ${1\over 59.1 sec}$ was found for a $1.3\,$M$_\odot$ pulsar   
model based on the tensor interaction equation of state. Surprisingly, the 35   
day period of HerX-1 fits the same formula almost exactly\footnote{Note that    
our argument differs from those of D'Alessandro and McCulloch (\cite{c8}), who  
rely on Shaham's (\cite{c17}) superfluid vortex theory in estimating that the   
angular momentum interplay between superfluid interior and the rigid crust is   
the most important mechanism determining the free precession frequency, and that
of Melatos \cite{c12}, who argues that $\omega_{fp}=\epsilon \omega_{rot}$      
``where $\epsilon$ is the non-hydrostatic ellipticity''.}. Our much larger      
dataset for the Crab pulsar strenghtens evidence for the 60 seconds period and
also suggests that this period is increasing with time at almost the same
rate as the rotation frequency of the pulsar. We also found that
different combinations of smaller data sets consistently produce a peak at
$\tilde \omega = {2\pi \over 59.93 s}$ (of course with a correspondingly smaller
signal to noise ratio). In Fig. 7 we plot the Fourier amplitude of the signal in
the expected frequency channel (assuming the frequency power law with $p=1$) as
a function of time together with errorbars (calculated from 10 neighbouring
amplitudes). Single data points are clearly very noisy, therefore, not much can
be said about the time dependence - in particular the glitch of July 1996
(Jodrell Bank Crab Pulsar Monthly Ephemeris \cite{c11}) did not leave a clear   
mark. The best linear fit to our data points gives an amplitude of $\Delta m(t) 
= (0.0045\pm 0.0006)-(0.00029\pm 0.00026) t$, where $t$ is expressed in years   
since 1996.0. 

\begin{figure}
   \resizebox{\hsize}{!}{\includegraphics{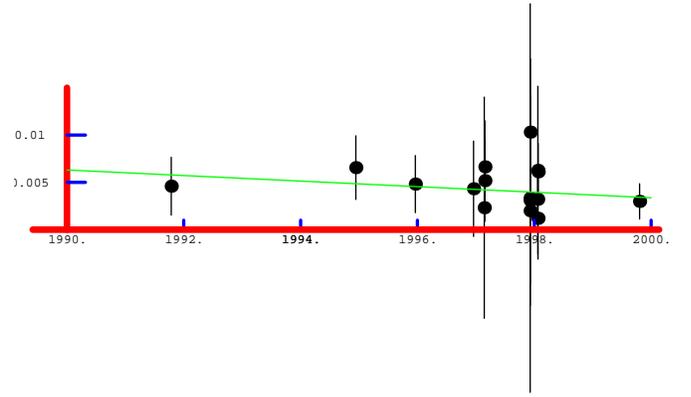}}
   \hfill
   \parbox[b]{55mm}{\caption[Fig7]{The spectral amplitude of the signal at the  
	expected frequency}}
   \label{fig:Fig7}
\end{figure} 

Does the pulsar free-precess? We presented all the evidence that we have and are
inclined to believe that it does point to an affirmative answer. Open questions 
remain such as: 
1) does the amplitude of the free precession really change
with time? and 2) on what time scale does the pulsar relax its internal stress? 
Observation of pulsar free precession can teach us much
about pulsar physics. As shown by \v Cade\v z, Gali\v ci\v c and Calvani        
\cite{c7}, the free precession model can quite sensitively (if something is     
known about the state of internal stress) distinguish between different neutron 
matter equations of state. Observing the change in free precession amplitude and
frequency can allow one to learn about external torques on the pulsar (jets may
produce them) and also about exchange of angular momentum between the crust and
the neutron superfluid. Many questions regarding the $1\over 60sec$ period      
remain open and, given the relevance of those questions to our understanding of 
pulsar physics and their interaction with neighbouring plasma, it would be      
important if other research groups could extend our analysis and observations.

%____________________________________________________________________________

\section{Acknowledgements}

We are grateful to many people who made these observations possible and who kept
our sometimes dwindling spirits up. The Padua Observatory is
among our staunchest supporters - in particular Massimo Calvani was supporting  
and stimulating our observations from the very beginning, Ulisse Munari granted 
us one of his nights at the 1.82m telescope and prof. Roberto Barbon was very   
helpful during our 1997 observations. We thank Boris Hrib for his help in the   
HP calibration lab and Jurij Kotar, Dick Manchester and David Nice for help with
the TEMPO program. The support of the technical staff of the Guillermo Haro     
observatory at Cananea is kindly acknowledged. It is a pleasure also to thank   
Jack Sulentic for his help in making the text more readable. This program was   
supported in part by the Ministry of Science and Technology of the 
Republic of Slovenia.

\end{document}